\documentclass[twoside]{article}

\newif\ifQIC
\QICfalse

\ifQIC
\usepackage{qic,epsfig}
\usepackage{qic}
\fi

\ifQIC
\else
\newcommand{\fcaption}{\caption}
\newcommand{\tcaption}{\caption}
\newtheorem{definition}{Definition}
\fi

\usepackage{cite}
\usepackage{url}
\usepackage{times}
\usepackage{graphicx}
\usepackage{color}
\usepackage{amsmath}
\usepackage{latexsym}
\newcommand{\ignore}[1]{ }

\newtheorem{observation}{Observation}
\newcommand{\B}{\bf }

\ifQIC
\else
\fi

\renewcommand{\thefootnote}{\fnsymbol{footnote}}  

\begin{document}

\ifQIC
\setlength{\textheight}{8.0truein}    
\else
\fi


\ifQIC
 \runninghead{FAST EQUIVALENCE-CHECKING FOR QUANTUM CIRCUITS}
             {SHIGERU YAMASHITA and IGOR MARKOV}

\normalsize\textlineskip
\thispagestyle{empty}
\setcounter{page}{1}

\copyrightheading{0}{0}{2003}{000--000}

\vspace*{0.88truein}

\alphfootnote

\fpage{1}

\centerline{\bf
FAST EQUIVALENCE-CHECKING FOR QUANTUM CIRCUITS}
\centerline{\footnotesize
SHIGERU YAMASHITA
}
\vspace*{0.015truein}
\centerline{\footnotesize\it Ritsumeikan University, 1-1-1 Noji Higashi}
\baselineskip=10pt
\centerline{\footnotesize\it Kusatsu, Shiga 525-8577, Japan}
\vspace*{10pt}
\centerline{\footnotesize
IGOR MARKOV}
\vspace*{0.015truein}
\centerline{\footnotesize\it University of Michigan, 2260 Hayward St.}
\baselineskip=10pt
\centerline{\footnotesize\it Ann Arbor, Michigan 48109, The United States of America}
\vspace*{0.225truein}
\publisher{(received date)}{(revised date)}

\vspace*{0.21truein}


\abstracts{
We perform formal verification of quantum circuits by integrating
several techniques specialized to particular classes of circuits. Our verification methodology is based on the new notion of a {\em reversible miter} that allows one to leverage
existing techniques for simplification of quantum circuits. For {\em
  reversible circuits} which arise as runtime bottlenecks of key
quantum algorithms, we develop several verification techniques and
empirically compare them. 
 We also combine existing quantum verification tools
 with the use of SAT-solvers. Experiments with circuits for Shor's number-factoring algorithm, containing thousands of gates, show improvements in efficiency by 3-4 orders of magnitude.
}{}{}

\vspace*{10pt}


\vspace*{3pt}
\communicate{to be filled by the Editorial}

\vspace*{1pt}\textlineskip    

\else

\title{Fast Equivalence-checking for Quantum Circuits}

\author{\normalsize
 \hspace*{-7mm}
 \begin{tabular}[t]{c@{\extracolsep{2em}}c}
  \large SHIGERU YAMASHITA& \large IGOR MARKOV \\
  \\
   Ritsumeikan University, 1-1-1 Noji Higashi & University of Michigan, 2260 Hayward St.\\
   Kusatsu, Shiga 525-8577, Japan & Ann Arbor, Michigan 48109, U.S.A.\\
   ger@cs.ritsumei.ac.jp & imarkov@eecs.umich.edu
\end{tabular}}
\maketitle

{\small\bf Abstract---
We perform formal verification of quantum circuits by integrating
several techniques specialized to particular classes of circuits. Our verification methodology is based on the new notion of a {\em reversible miter} that allows one to leverage
existing techniques for simplification of quantum circuits. For {\em
  reversible circuits} which arise as runtime bottlenecks of key
quantum algorithms, we develop several verification techniques and
empirically compare them. 
 We also combine existing quantum verification tools
 with the use of SAT-solvers. Experiments with circuits for Shor's number-factoring algorithm, containing thousands of gates, show improvements in efficiency by 3-4 orders of magnitude.
}
\fi

\section{Introduction}

 In August 2009 ``{\em researchers at the [US] National Institute of Standards and Technology ... demonstrated continuous quantum operations using a trapped-ion processor}''~\cite{EETimes09} that maintained quantum bits in
 hyperfine states of beryllium ions for up to 15 seconds at
 a time. An implementation account of NIST's quantum 
 processor~\cite{NIST2q,NIST2qS} shows that the design of even two-qubit circuits relies on software tools,
 similar in spirit to logic-synthesis and optimization 
 tools used today to design digital logic circuits. 
   A large-scale architecture for quantum computing proposed
   in June 2009 suspends linear ion crystals in an anharmonic trap~\cite{NewTrap09}. A concrete design that implements this architecture provisions for 100 ytterbium-based logical qubits and 20 additional ions for laser cooling. 
 Additional applications of quantum circuits --- commercial quantum communications and cryptography --- have so far relied on quantum-optical implementations, where qubits are carried by photons over great distances. In August 2009, Siemens and  IdQuantique announced commercial availability in Europe of
   quantumly-secure communication through unused standard fiber-optic cables (``dark fiber'') \cite{Siemens09}.
Several photonic realizations of Shor's number-factoring algorithm have
been reported since 2007~\cite{Lanyon2007,Lu2007}, including a single-chip circuit~\cite{Politi09}.


 Quantum circuits often operate on quantum states that contain exponentially large superpositions, making quantum simulation, as well as circuit design and analysis on conventional computers very challenging. To this end, a layered software architecture for quantum computing design tools was outlined in \cite{Svore06}. This work focuses on one such task --- verifying the results of quantum circuit transforms, e.g., adaptations of technology-independent quantum circuits to linear device architectures, such as ion traps
 \cite{FowlerEtAl2004,Maslov07}. In other words, given a circuit that is known to be correct, one seeks to prove that a new circuit optimized for a given physical technology
 is equivalent to the original circuit.

Past research in equivalence-checking for quantum circuits developed computational techniques based on
Binary Decision Diagrams (BDDs)~\cite{MT06,VMH03,VMH07}.
These techniques can represent some exponentially large complex-valued vectors and matrices using compact graphs.
Quantum operations are then modeled by graph algorithms
whose complexity scales with graph size rather than with
the size of superpositions or the amount of entanglement present.
  However, these algorithms are much slower than those for equivalence-checking of conventional digital logic and
  do not scale to useful instances of Shor's algorithm.

 An important observation is that a typical quantum algorithm consists of heterogeneous modules~\cite{Nielsen2000}
that favor different computational techniques for
equivalence-checking. This motivates the development of a new verification methodology that invokes
the most appropriate technique for each module type and assembles the results.
Our methodology relies on a new concept, introduced in Sec.~\ref{sec:RevMiter} and called {\em a reversible miter} --- a natural counterpart of {\it miter circuits} used
in equivalence-checking of digital electronic circuits.
In conjunction with existing techniques for iterative circuit
simplification~\cite{ger2002-1,MaslovDMN07,PrasadEtAl2007},
reversible miters can drastically reduce the size and complexity of circuits under verification, especially when such circuits bear some structural resemblance (which is often the case when adapting textbook circuits to specific quantum-computing architectures).

In Sec.~\ref{sec:adaptive} we develop an high-performance equivalence-checking for quantum circuits. Our method
is {\it adaptive} in the sense that it utilizes multiple techniques appropriate for different classes of quantum circuit modules. In this context, we study {\it reversible circuits} which are a subset of quantum circuits that map conventional 0-1 bit-strings into other such bit-strings.
 In particular, the largest module in Shor's number-factoring
 algorithm~\cite{Shor97} --- {\it modular exponentiation} --- is
 implemented as a reversible circuit \cite{MeterI2005}
 (acting on entangled quantum states), exceeds all other modules asymptotically in size, and thus requires
 most attention of CAD tools.
 To verify such logic modules, we adapt conventional state-of-the-art techniques \cite{MishchenkoEtAl2006,ABC} in several ways, and significantly scale up quantum equivalence checking. Empirical comparisons
 in Sec.~\ref{sec:ConvVerify} confirm that
 properties of reversible circuits can enable much faster
 SAT-based equivalence-checking.
 However, conventional techniques cannot be applied to,
 e.g., the {\it Quantum Fourier Transform (QFT)}.
 Therefore, we also study equivalence-checking of circuits with
 non-conventional gates (we call these circuits {\em properly-quantum}), and the integration of heterogeneous
 techniques.

  Our contributions can be summarized as follows.
\begin{itemize}
\item {\it Reversible miters} for equivalence-checking
  of quantum circuits, and their integration with circuit simplification.
\item The use of SAT-based equivalence checking
      and its integration with BDD-based techniques.
\item Adaptive equivalence-checking
   for quantum circuits that integrates reversible miters,
   circuit simplification, as well as SAT- and BDD-based techniques.
\end{itemize}




\section{Notation and Preliminaries}\label{sec:QC}
\ignore{
We now provide background in quantum computation \cite{Nielsen2000}, necessary to understand our work.

\noindent {\bf Qubits.}
While a {\it bit} is a fundamental unit of (conventional) information, {\em quantum information} is expressed in terms of {\it quantum bits}, or {\it qubits} for short.  A qubit is a mathematical abstraction of a {\it quantum state} such as
nuclear spin of an atom.  {\em Basis states} of a qubit are labeled $|0>$ and $|1>$.  A qubit can assume any complex-valued linear combination of basis states
$\alpha\left|0\right>+\beta\left|1\right>$ with $|\alpha|^2 + |\beta|^2 = 1$,
i.e., a norm-1 vector
 $     \left(
        \begin{smallmatrix}
        \alpha \\
        \beta \\
        \end{smallmatrix}
      \right)$.
}

\begin{figure}[tbp]
\hspace*{-1mm}
\begin{center}
\begin{minipage}{8cm}
\begin{center}
\resizebox{5cm}{!}{
\includegraphics*{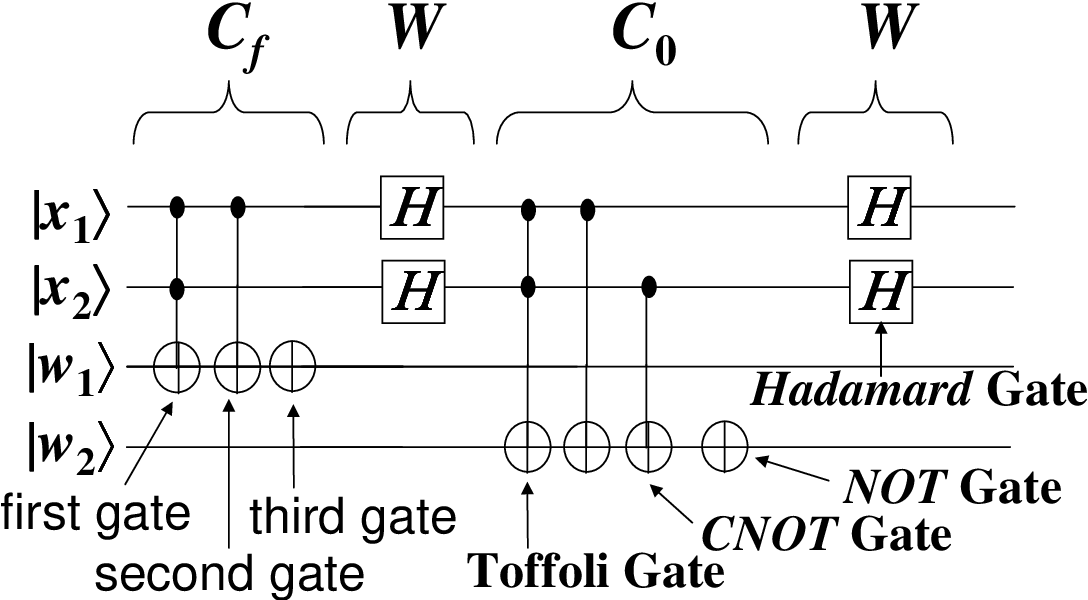}
}
\fcaption{A properly-quantum circuit (iteration of Grover algorithm).}\label{fig:GS}
\end{center}
\end{minipage}
\end{center}

\hspace*{-1mm}
\begin{center}
\hspace{-3mm}
\includegraphics*[width=3cm]{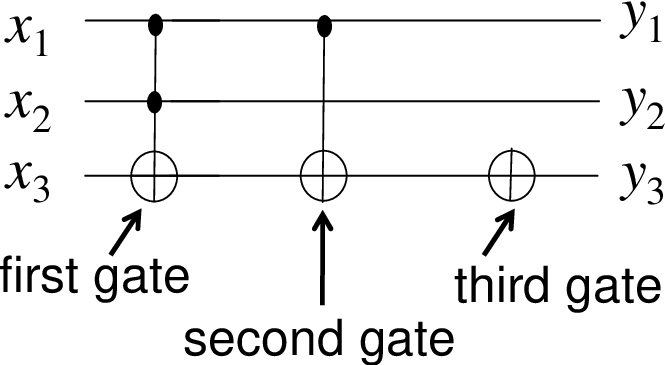}
\hspace{2mm}
\includegraphics*[width=4.2cm]{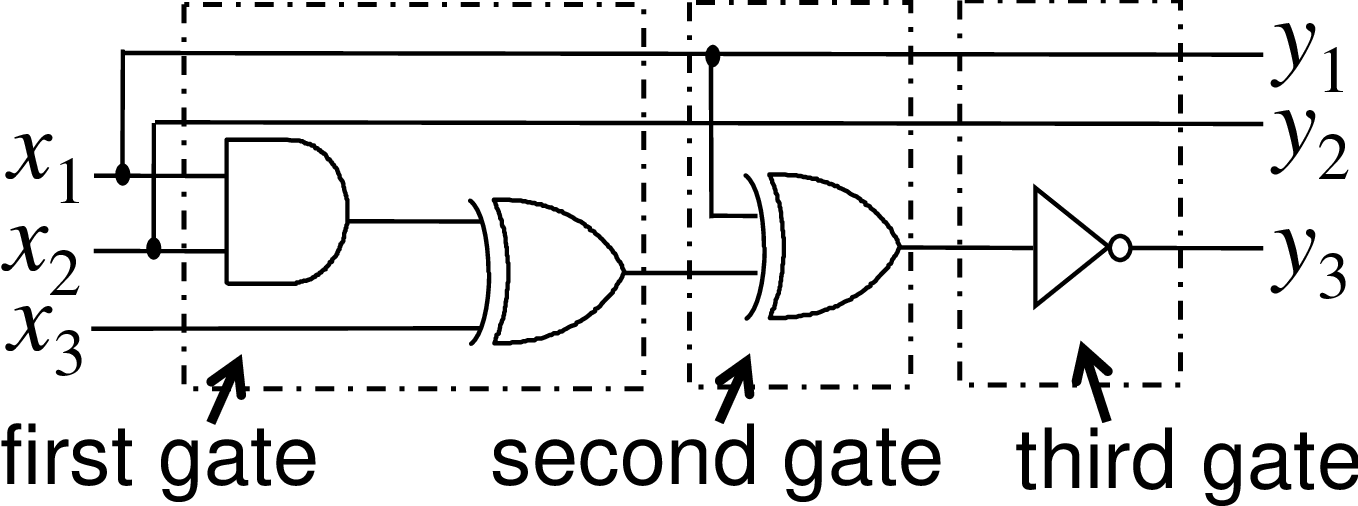}
\fcaption{A reversible circuit and its irreversible realization.}\label{fig:Circuit1} \label{fig:ConvCirc}
\end{center}
\end{figure}

Recall that, when acting on conventional bits, gates NOT, CNOT and TOFFOLI can be implemented using NOT, XOR and AND gates as shown in Fig. \ref{fig:ConvCirc}. In the quantum case, they
exchange basis states, which is why their matrices contain
only 0s and 1s. As these gates obey the same algebraic rules
 in both cases, we term them {\it conventional gates}.
In comparison, the matrix of the Hadamard gate contains $1/\sqrt{2}$, and its functionality cannot be expressed in
Boolean logic. Therefore we call such gates {\bf properly-quantum}. Each properly-quantum gate maps at least one 0-1 input combination (basis state) to a quantum superposition of more than one basis state.
Circuits that include properly-quantum gates are also called
properly-quantum. An example is given in Fig.~\ref{fig:GS}.
As we show below, many reversible circuits without properly-quantum gates can be
verified relatively easily in practice using a state-of-the-art equivalence-checking tools for conventional logic circuits
based on solving instances of Boolean SATisfiability. Modern SAT-solvers exploit structure in application-derived instances, and modern equivalence-checkers automatically identify and exploit similarities in the circuits whose equivalence is checked.

\ignore{
Measurement-free quantum computation is reversible in nature \cite{Nielsen2000},
 therefore quantum circuits map their input configurations to output configurations one-to-one, and this property is also observed locally for every gate and sub-circuit \cite{Nielsen2000}.
}

Many quantum algorithms contain large, application-specific sections dedicated to the computation of Boolean functions.
In order to embed conventional computation into the quantum domain, it must be made reversible, and standard procedures exist for such transformations \cite{Nielsen2000}. The
resulting circuits do not create entanglement,
but can be applied to superposition superposition states. Leveraging this quantum parallelism in useful applications
is difficult, but can be illustrated by Shor's polynomial-time
algorithm for number-factoring \cite{Shor97,Nielsen2000}.
This algorithm is dominated by a reversible module that performs modular exponentiation \cite{MeterI2005} before the Quantum Fourier Transform (QFT).
We call such circuits without properly-quantum gates specifically {\bf reversible circuits} in this paper.
A gate library used for reversible circuits is {\em universal}
iff it can express any (conventional) reversible transformation by combining multiple copies of gates involved.
The most common such gate library consists of NOT, CNOT, and Toffoli gates.
%
Since the algebraic properties of the gates in reversible circuits do
not involve quantum phenomena,  we can calculate the logic functions
realized at each point in a circuit, as is normally done in
conventional logic synthesis and verification. For example, we can calculate the function
at wire $x_3$ of the circuit shown in Fig.~\ref{fig:Circuit1}
 after the third gate as $y_3 = x_3 \oplus x_1 x_2 \oplus x_1 \oplus 1$.

 \ignore{ In contrast, output qubits of properly-quantum circuits are not independent from each other because quantum information is carried by cross-correlations. To capture these correlations, the function implemented by a properly-quantum circuit is represented by a large complex-valued matrix rather than a Boolean formula.
 }


\section{Reversible Miters}\label{sec:RevMiter}

To check the equivalence of two combinatorial digital logic
circuits, $C_1$ and $C_2$, one checks if
the conventional {\it miter circuit}~\cite{MishchenkoEtAl2006}
shown in the left-hand side of Fig.~\ref{fig:ConvMiter}
implements the constant-0 function.
In other words, every pair of outputs are XOR'ed,
all XOR-outputs are OR'ed together, and the resulting Circuit-SAT instance is converted to CNF-SAT using known techniques (a number of optimized reductions have been proposed
recently with large circuits in mind).
Conventional miters can be constructed for reversible circuits
by treating them as AND/OR/NOT circuits, except that such miters will not be reversible.
Therefore, we introduce {\it reversible miters} which can
handle reversible and properly-quantum circuits and
can benefit from simplification of reversible circuits~\cite{ger2002-1,PrasadEtAl2007,MaslovDMN07}.

%
%

\subsection{Properties of Quantum Circuits}\label{subsec:RevMiter}



 Observe that for quantum or reversible circuits $C_1$ and $C_2$,
  the concatenated circuit $C_1 \cdot C_2$ is of the same kind.
Such circuits can also be structurally reversed.

 \begin{observation}
Given a quantum (or reversible) circuit $C = g_1 \cdot g_2 \cdot \cdots
\cdot g_{k}$ where $g_i$ is a gate,
its copy where all gates are inverted and put in the reverse order,
i.e., $g_k^{-1} \cdot \cdots \cdot g_2^{-1} \cdot g_{1}^{-1}$,
 implements the inverse transformation to what $C$ implements. We
 therefore denote it by $C^{-1}$.
 \end{observation}

\begin{figure}[t]
 \begin{center}
\resizebox{4cm}{!}{
 \includegraphics*{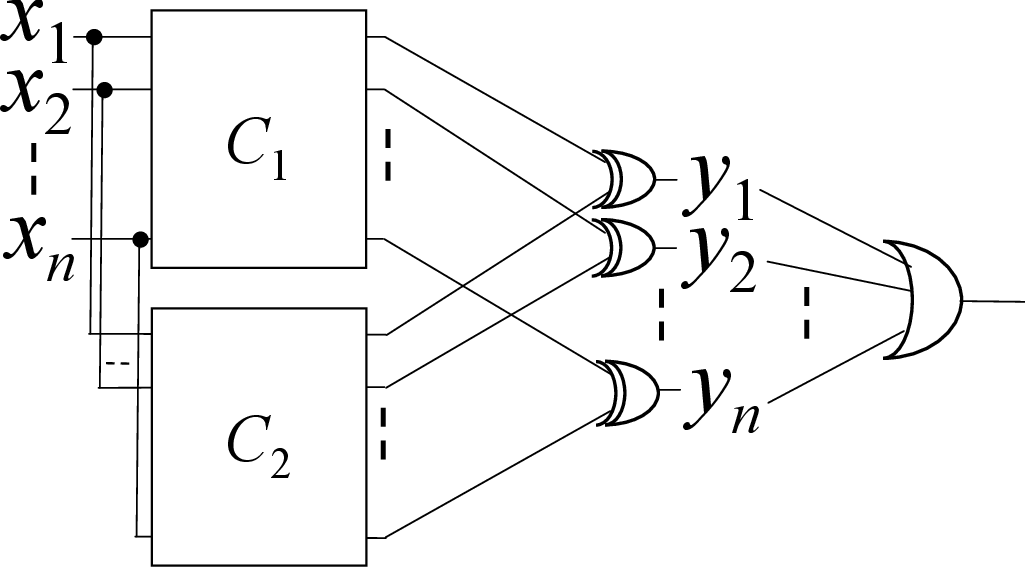}
 }\ \ \ \
 \resizebox{3.77cm}{!}{
 \includegraphics*{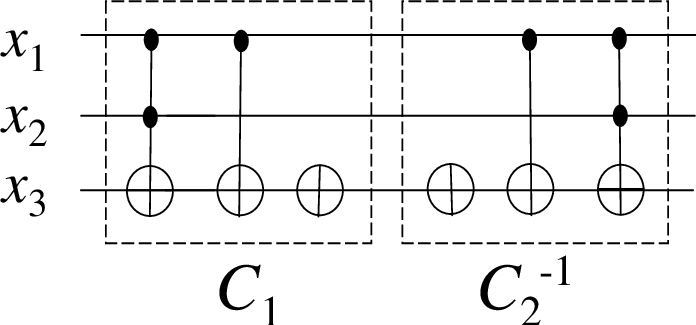}
 }
 \fcaption{Miter circuits: conventional and reversible. }\label{fig:Reversible Miter} \label{fig:ConvMiter}
 \end{center}
 \end{figure}

For example, for a circuit $C$ shown in the left-hand side of
Fig.~\ref{fig:Circuit1}, the circuit
$C \cdot C^{-1}$ is given in the right-hand side of Fig.~\ref{fig:Reversible Miter}.
Note that  NOT, CNOT, and Toffoli gates are their own inverses
(which explains their choice as library gates).
The circuit $C \cdot C^{-1}$ is equivalent to an {\it empty circuit}.
This can be confirmed by iteratively cancelling out pairs of
mutually-inverse adjacent gates.
Namely, in the right-hand side of Fig.~\ref{fig:Reversible
  Miter}, the third and the fourth gates can be removed at once.
Then, the second and the fifth gates, followed by the first and the
last gates.  This observation motivates our new notion of {\em
  reversible miters}.


\subsection{Reversible Miter Circuits}

\begin{definition}
Given two quantum (or reversible) circuits $C_1$ and $C_2$, their {\em
  reversible miter} is defined to be one of the
following circuits: $C_1\cdot{C_2}^{-1}$, ${C_2}^{-1} \cdot
{C_1}$, ${C_2} \cdot {C_1}^{-1}$, ${C_1}^{-1} \cdot {C_2}$.
\end{definition}

 In particular, for conventional miters one needs to check
 that the output functions implement the constant 0 function, whereas
 for reversible miters one checks that each output bit is equivalent to
 a corresponding input bit. Namely, $C_1$ and $C_2$ are functionally
 equivalent if and only if all of their reversible miters implement the identity
 transformation. In particular, if one miter implements the identity,
 then so do the remaining miters.
If $C_1=C_2$, then straightforward circuit
simplification~\cite{ger2002-1,PrasadEtAl2007,MaslovDMN07}
cancels out all gates, resulting in an empty circuit. Some of the variant miters enable more cancellations than others, e.g., if $C_1$ and $C_2$ differ only in their first segments, $C_2 \cdot C_1^{-1}$ exhibits
many gate cancellations.

Reversible miters speed up equivalence-checking by exploiting similarities in circuits by two
 distinct mechanisms.

\subsubsection{Local Simplification of Reversible Miters}\label{sec:ReversibleMiterSimplification}

   When two conventional circuits end with identical gate sequences, one cannot
cancel out these sequences because of observability don't-cares introduced by
them. However, reversible circuits do not experience don't-cares, and identical
suffixes always cancel out.  Note that a reversible miter $C_1 \cdot C_2^{-1}$
places the last gate of $C_1$ next to the last gate of $C_2$. If these two
gates cancel out, the second-to-last gates from $C_1$ and $C_2$ become
adjacent, etc. Thus, no search is required to identify these gate
cancellations, and they can be performed one at a time. Even if the last
two gates are different, it may be possible to cancel out second-to-last
gates, as long as the last and second-to-last gates do not act
on the same (qu)bit lines.
These are special cases of much more general {\it local simplifications} discussed
in \cite{ger2002-1,PrasadEtAl2007,MaslovDMN07}.  If $C_1$ and $C_2$ are
identical, an empty circuit will result, but this outcome is also possible when local simplifications can prove equivalence of two structurally different circuits.
A systematic procedure for applying simplifications was introduced in \cite{ger2002-1}.
Local simplifications in reversible circuits are particularly easy to perform,
are fast and do not consume much memory \cite{PrasadEtAl2007,MaslovDMN07}.
   In our experiments, even the simplest simplification rules can
   dramatically simplify reversible miters. More sophisticated simplifications
   from \cite{ger2002-1,PrasadEtAl2007,MaslovDMN07}
   provide an additional boost.

\begin{figure}[t]
\begin{center}
 \includegraphics*[width=3.5cm]{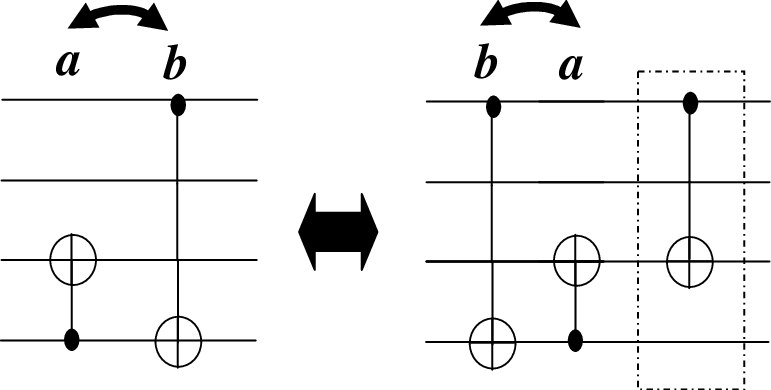}\ \ \
 \includegraphics*[width=3.5cm]{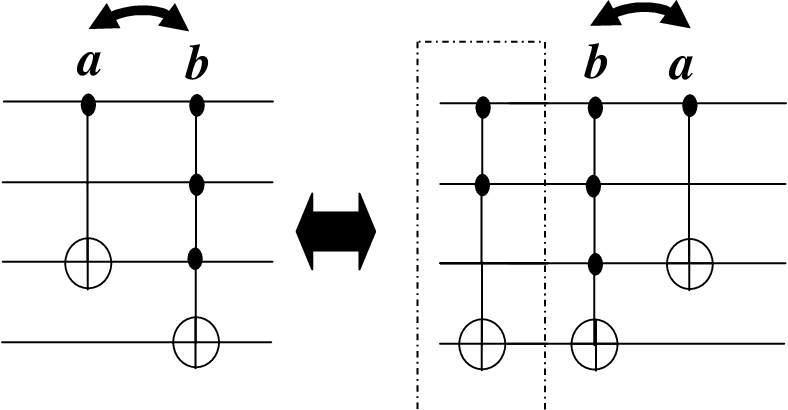}\ \\
\fcaption{Gate swap (complicated case).}\label{fig:rule3}
\end{center}
\end{figure}

   We experimented with the following simplification procedure.
   In a miter circuit, consider one gate at a time, search for
   a matching inverse, and try to move them together to facilitate cancellation. Any two gates can be swapped if they do not act on the same (qu)bit lines.
   Two adjacent NOT, CNOT or Toffoli gates can be swapped if the control bit of one gate is not the target bit of the other gate (same for properly-quantum controlled-$U$ gates).
%
   A more sophisticated swapping rule (for NOT, CNOT, and Toffoli gates)
   is illustrated in Fig.~\ref{fig:rule3}.

\begin{figure}[tp]
  \begin{center}
  \begin{minipage}{8cm}
  \begin{center}
  \resizebox{7.5cm}{!}{
  \includegraphics*{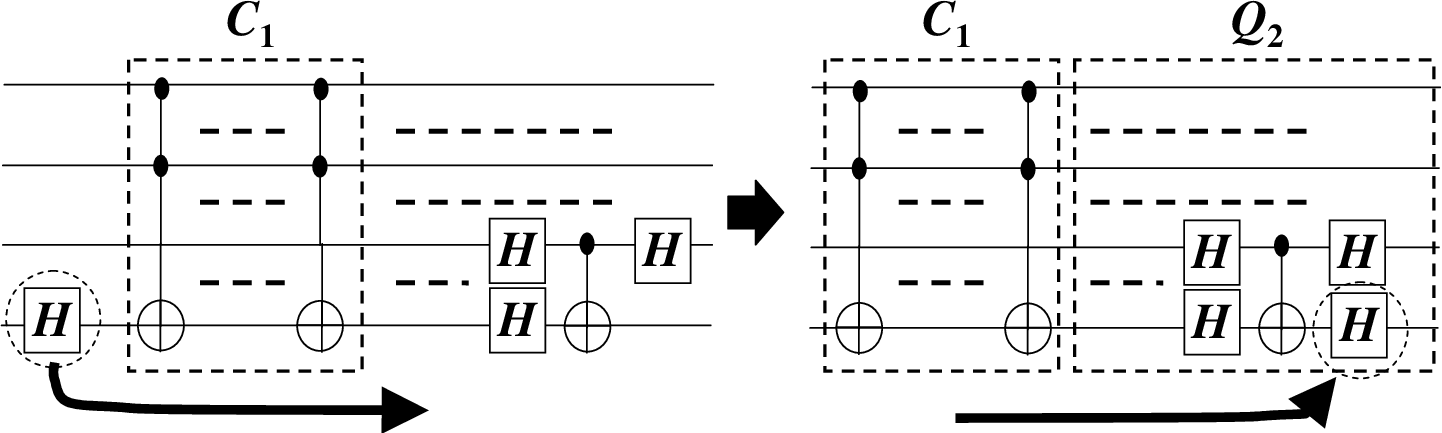}
  }
  \fcaption{Transforming a miter circuit after simplification.}\label{fig:extension} \label{fig:Circular}
  \end{center}
  \end{minipage}
  \end{center}
 \end{figure}

   In our procedure, for the purposes of equivalence-checking, we temporarily
consider the miter circuit to be ``circular'' by connecting its outputs to its
inputs.
Namely, we allow moving the first gate to the end of the circuit, as illustrated in Fig.~\ref{fig:Circular}.
 This transformation does not change the equivalence of the entire
 circuit to the identity.  In other words, if  $g_1 \cdot g_2 \cdot \cdots \cdot
 g_{k-1} \cdot g_k=I$ (Identity), then $g_1^{-1} \cdot g_1 \cdot g_2 \cdot
 \cdots \cdot g_{k-1} \cdot g_k \cdot g_1 = g_1^{-1} \cdot I \cdot
g_1 = g_1^{-1} \cdot g_1 = I$. Therefore, to check equivalence between $g_1 \cdot
 g_2 \cdot \cdots \cdot g_{k-1} \cdot g_k$ and $I$ is the same as to check
 equivalence between $g_2 \cdot \cdots \cdot g_{k} \cdot g_{1} $ and
 $I$.

\begin{figure}[tbp]
 \begin{center}
\resizebox{4cm}{!}{
 \includegraphics*{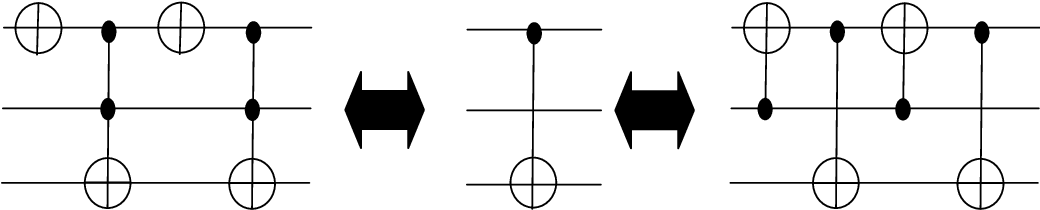}
 }\ \ \ \
 \resizebox{3.77cm}{!}{
 \includegraphics*{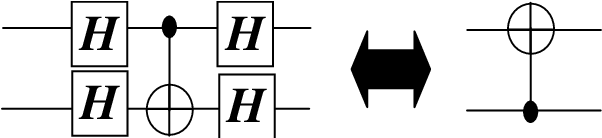}
 }
 \fcaption{Equivalent circuit templates. }\label{fig:templates}
 \end{center}
 \end{figure}

  A variety of circuit-equivalence templates can be
used with the above simplification
procedure~\cite{MaslovDMN07,PrasadEtAl2007,Tucci2004}
to shrink the miter circuit. Such templates are known for
both reversible and properly-quantum gates as shown
in Fig.~\ref{fig:templates}. For example, the transformation
illustrated
in Fig.~\ref{fig:Circular} enables further simplification through
the equivalence in Fig.~\ref{fig:templates} on the right.

\subsubsection{Simplification of Canonical Forms}

Iterative circuit simplification is not guaranteed to reduce
$C_1\cdot C_2^{-1}$ to the empty circuit in polynomial time
when such a simplification is possible. Finding a short
simplification may be time-consuming.
Yet, when constructing canonical forms (ROBDDs or QuIDDs) of reversible miters, a different kind of simplification may occur. Suppose that $C_1$ and $C_2$ end with functionally-equivalent but structurally distinct suffixes that do not admit local simplifications --- an example is given in \cite{PrasadEtAl2007}. In other words $C_1=A_1\cdot
B_1$ and $C_2=A_2\cdot B_2$ where $B_1\approx B_2$.
Then $C_1\cdot C_2^{-1}=A_1\cdot B_1 \cdot B_2^{-1} \cdot
A_2^{-1} \approx A_1 \cdot A_2^{-1}$.

   As we traverse the miter $C_1\cdot C_2^{-1}$, adding one gate at a time to the decision diagram (DD), the size of the intermediate DDs depends only on the transformation implemented by the current circuit prefix, i.e., the functions of the intermediate wires. The intermediate DD for
   $ A_1\cdot B_1 \cdot B_2^{-1}$ can be smaller than that for
   $ A_1\cdot B_1$ if $ A_1\cdot B_1 \cdot B_2^{-1} \approx A_1$.
   This phenomenon was observed in our experiments.



\section{Equivalence-checking for Quantum Circuits}\label{sec:adaptive}

We now introduce equivalence-checking of quantum circuits based on several techniques appropriate for different classes of quantum circuits. The first class contains reversible circuits that arise as key modules in quantum algorithms.

\subsection{Equivalence-checking for Reversible Circuits}\label{sec:ConvVerify}


To check the equivalence of two reversible circuits, $C_1$ and
$C_2$, one can pursue two strategies. The first strategy is to
check that the conventional miter implements
the constant 0 function.  A conventional miter can also be
applied to reversible circuits as explained below.  The second
strategy is to represent the transformations performed by $C_1$ and $C_2$
in a canonical form which supports efficient equivalence-checking.

The latter strategy may use binary-decision diagrams (BDDs), such as ROBDDs, and QuIDDs \cite{VMH03} or QMDDs \cite{MT06}. The former can be implemented with either decision diagrams or Boolean Satisfiability solvers by reducing Circuit-SAT to
CNF-SAT. In particular, for conventional miters one needs to check
 that the output functions implement the constant 0 function.  In addition to
the basic SAT or BDD-based approaches, finding equivalent signals in two
circuits is often very helpful~\cite{MishchenkoEtAl2006}.  Such techniques
appear useful for reversible circuits as well, as shown in our experiments. Relevant computational engines are discussed next.

\vspace*{0.1cm}
\noindent{\bf ROBDD.}
Calculate the output functions of miter circuits, using ROBDD as the primary data structure. This technique cannot handle properly quantum circuits.

\vspace*{0.1cm}
\noindent{\bf QuIDD.}
   Build functional representations of given circuits $C_1$
   and $C_2$, and check if the results are identical.
   In particular, QuIDDPro \cite{VMH03,VMH07} builds multi-terminal decision diagrams called QuIDDs that can capture properly-quantum circuits.

\vspace*{0.1cm}
\noindent{\bf SAT.}
  Given two reversible circuits, construct a CNF-SAT formula that is satisfied only by those input combinations for which the two circuits produce different outputs.
 Then use a contemporary SAT solver \cite{{MiniSAT}} to check satisfiability.\footnote{
 Recall that NP-completeness relates to worst-case complexity
 and does not prevent fast solution of many application-derived
 SAT instances. In industrial applications, modern SAT solvers can often resolve CNF-SAT instances with hundreds of thousands variables in several hours, although small hard instances are also known.}
We construct a CNF formula as follows.
First we add a set of clauses for each gate in the miter circuit. The clauses should be
satisfied only with the variable assignments that are consistent
with the reversible gate. The readers familiar with SAT-based
equivalence-checking can think of a CNOT gate as an XOR gate with a
bypass wire, and of a Toffoli gate as an XOR, AND and a bypass. More
efficient clause generation is illustrated below for a Toffoli gate
whose control bits are $x_1$ and $x_2$, and target bit is $x_3$.
Since the Toffoli gate does not modify two of its inputs, there is
no need for separate output variables. We introduce only one new
variable $y_1$ for the target bit. Then logical consistency is given
by the condition $ y_1 = (x_1 \cdot x_2) \oplus x_3$ which can be
expressed by the following six clauses.
\begin{itemize}
  \item
   Case $x_1=0$ or $x_2=0$. 
  Clauses: $(x_1 + \overline{x_3} + y_1) \cdot (x_1 + x_3 + \overline{y_1})
  \cdot (x_2 + \overline{x_3} + y_1) \cdot (x_2 + x_3 +
  \overline{y_1})$.
  \item Case $x_1=x_2=1$. 
  Clauses:
  $(\overline{x_1} + \overline{x_2} + x_3 + y_1) \cdot (\overline{x_1}
  + \overline{x_2} + \overline{x_3} + \overline{y_1})$.
\end{itemize}

In the next step,
we add a set of clauses that are satisfied only by those variable
combinations where some circuit output differs from the respective
circuit input.

Here we can reuse some of the $y$ variables introduced earlier.
Let such a new variable corresponding to the $i$-th primary output be
 $y_{O_i}$. (If there is no target bit on the $i$-th bit-line, we do
  not introduce a new variable for the $i$-th primary output, i.e.,
 it is obvious that the input and the output functions on the
 $i$-th bit-line are the same, and thus we do not add the following
  clauses.)
 We introduce a new variable $z_i$ to express the functional
 consistency of the $i$ bit-line. Namely, we consider
  that $z_i$ becomes 1 only when $x_i \neq y_{O_i}$.
For this condition, we add the following clauses.
\begin{itemize}
  \item
   Case $z_i=0$.
   Clauses: $(z_i + x_i + \overline{y_{O_i}}) \cdot  (z_i + \overline{x_i} + y_{O_i})$.
  \item
   Case $z_i=1$.
   Clauses: $ (\overline{z_i} + x_i + y_{O_i}) \cdot  (\overline{z_i}
   + \overline{x_i} + \overline{y_{O_i}})$.
\end{itemize}
Finally we add $(z_1 + z_2 + \cdots + z_n)$ where $n$ is the
number of bit-lines of the circuits.
Since $z_i=1$ mens that the input and the output functions on
the $i$-th bit-line are different, the two circuits are different
when $(z_1 + z_2 + \cdots + z_n)$ is
satisfied. Therefore, the above construction generates a SAT formula
that is satisfied only by those input combinations for which the
corresponding outputs of two circuits produce different values.
A CNF-SAT formula constructed for a miter grows linearly with the
size of the miter. A key advantage of reversible miters is that they
can be significantly smaller, due to gate cancellations and other
circuit simplifications.

\vspace*{0.1cm}
\noindent{\bf State-of-the-art Combinational Equivalence Checking.}
   SAT-based techniques can be dramatically improved through synergies with randomized functional simulation and through
   identifying intermediate equivalences.  By hashing the results of random simulation, one finds candidate equivalent wires. If $w_1$ and $w_2$ are {\em not} equivalent, the counterexample returned by SAT is used to refine the results of functional simulation and often distinguishes other seemingly-equivalent pairs of wires. Once intermediate wires $w_1$ and $w_2$ are proven equivalent, all downstream gates are reconnected to $w_1$, and $w_2$ can be excluded from the SAT instance (along with some of its upstream gates). 
   If potentially equivalent wires are selected in a topological order from the inputs, the impact of multiple circuit restructuring steps accumulates, until all output wire are proven equivalent or until an input combination is found that disproves the equivalence of outputs.
    
The state-of-the-art implementation of these techniques
found in the Berkeley ABC system~\cite{ABC} (the ``cec'' command) features incremental SAT-solving and {\it fraiging} ---  a fast circuit-simplification technique based on hashing \cite{MishchenkoEtAl2006}. To use ABC, we construct a conventional (irreversible) circuit from a reversible circuit as shown in Fig.~\ref{fig:ConvCirc}.

   The impact of random-simulation techniques on SAT-based equivalence-checking can be illustrated
   by the example of multiplier circuits, which
   are known to confound both BDD-based and SAT-based
   computations. The case of equivalent multipliers is particularly difficult because it cannot be quickly concluded by finding (perhaps, by luck) input combinations
   that disprove the equivalence. However, if the two given multipliers are structurally similar and include many equivalent wires, then global equivalence can
   be proven quickly through a series of lemmata. Empirical data in Table \ref{tab:exp4} shows that on a 6-bit multiplier CEC outperforms by far BDD-based and SAT-only methods.
   

  Common benchmarks for reversible circuit synthesis
  can be verified in milliseconds by the above techniques.
  Therefore, we focus on scalable blocks of standard quantum algorithms, whose optimization and equivalence-checking are
 critical to the success of quantum computers being designed today.
More concretely, we performed experiments with $n$-bit {\it
linear-nearest-neighbor (LNN)} CNOT gate circuits, a reversible ripple-carry adder circuit proposed in~\cite{Cuccaro2004}, {\it mesh} circuits~\cite{FowlerEtAl2004} and reversible multipliers. Given a (qu)bit ordering, a linear-nearest-neighbor (LNN) CNOT gate circuit is a
circuit which realizes the functionality of a CNOT gate with target and control
bits $k$ bits apart, by using only LNN gates (gates that operate only on
adjacent qubits). Studies of LNN architectures are important because
several promising implementations of quantum computation require
the LNN architecture (also called the {\em spin-chain} architecture in
the physics literature) and allow only adjacent qubits to interact directly.
Thus, standard quantum circuits must be adapted to such architectures and modified to use only LNN gates.  Specific transformations and LNN circuits have been developed \cite{FowlerEtAl2004,Maslov07}.
The overhead of the LNN architecture in terms of the number of gates is
often limited by a small factor (3-5). Such physical-synthesis
optimization motivates the need for equivalence-checking against the original, non-LNN versions.
Using important components of Shor's algorithm~\cite{Nielsen2000,FowlerEtAl2004}
 --- adders, meshes and multipliers --- we build three types
 of equivalence-checking instances.

\noindent {\bf Same.} Two equivalent circuits.

\noindent {\bf Different~1.}
Add ten random Toffoli gates at the end.

\noindent {\bf Different~2.}
Add ten random Toffoli gates at the beginning.

Our empirical data for CNOT, adder and mesh circuits exhibits essentially the
same trends. Hence we report results only for adders in Table~\ref{tab:exp2}.
All runtimes are for a Linux system with a 2.40GHz
Intel\textregistered\ Xeon\texttrademark\ CPU with 1GB RAM.

\begin{table}[tb]
\begin{center}
\tcaption{Adder verification performed by several techniques.}\label{tab:exp2}
\scalebox{1.0}{
\begin{tabular}{|c|c|c|c|c|c|c|c|} \hline
Case &  $n$  & $\sharp$qubits  & $\sharp$gates & SAT  & QuIDD &  BDD & cec   \\ \hline
Same & 32  & 66    & 280    & 0.65    & 20.10   & \B 0.03 & \B 0.19 \\ \cline{2-8}
     & 64  & 130   & 568   & 2.91   & 115.85    & \B 0.11 & \B 0.23 \\ \cline{2-8}
     & 128 & 258   & 1144   & 11.71       & 771.20     &  \B 0.52 & \B 0.31 \\ \hline \hline
Diff.~1 & 32 & 66    & 290    & 1.00    & 31.93   & \B 0.04 & \B 0.02 \\ \cline{2-8}
 & 64  & 130   & 578   & 5.16   & 212.57    & \B 0.25 & \B 0.26 \\ \cline{2-8}
 & 128 & 258   & 1154   & 15.25       & $>$ 1,000     &  1.67 & \B 0.38 \\ \hline \hline
Diff.~2 &  32  & 66   & 290   & 1.09    & 40.40   &  0.09 & \B 0.02 \\ \cline{2-8}
&  64  & 130   & 578   & 10.98   & 318.62    & 0.76 & \B 0.03 \\ \cline{2-8}
&  128 & 258   & 1154   & 22.72       & $>$ 1,000     &  9.88 & \B 0.03 \\ \hline
\end{tabular}
}
\end{center}
\begin{center}
\tcaption{Multiplier verification performed by several techniques.}\label{tab:exp4}
\scalebox{1.0}{
\begin{tabular}{|c|c|c|c|c|c|c|c|} \hline
 &  $n$  & $\sharp$qubits  & $\sharp$gates  & SAT  & QuIDD &  BDD & cec  \\ \hline
Same & 4 &  20     & 166    & 1.86  & 50.45   & 0.09  & \B 0.00 \\ \cline{2-8}
     & 6  & 30     & 411  & 392.74    & $>$ 1,000   & 39.19 & \B 0.01 \\ \hline  \hline
Diff.~1 &  4 & 20  & 176    & \B 0.02    & 72.84   & \B 0.01 & \B 0.01 \\ \cline{2-8}
         & 6 & 30  & 421    & 0.11    &  $>$ 1,000    & \B 0.03 & \B 0.02 \\ \hline \hline
Diff.~2 & 4  & 20  & 176   & \B 0.02    & 95.94   & \B 0.01  & \B 0.02 \\ \cline{2-8}
        & 6  & 30  & 421   & 0.17    &  $>$ 1,000    & \B 0.01  & \B 0.02 \\ \hline
\end{tabular}
}
\end{center}
\end{table}

We implemented $n$-bit reversible multipliers using $5n$ bit-lines, including $2n$ bits for two inputs,
$2n$ bits for the results, and $n$ ancillae.
E.g., the line $n=6$ in the tables deals with $30$-bit circuits. The $n$-bit adder circuit proposed in~\cite{Cuccaro2004} uses $2n+2$ qubits.
Thus, the third column in Tables~\ref{tab:exp2} and \ref{tab:exp4} shows the number of qubits in each circuit.
The forth column shows the number of gates in each circuit.
All methods other than ``cec''  timed out for $n=8$, requiring more than 1,000s.

\subsection{Checking Properly-Quantum Circuits}\label{sec:extension}

  In this section we show that our proposed techniques can handle properly-quantum gates, but remain compatible with fast special-case methods.

\subsubsection{Utility of Reversible Miters}
Earlier sections focused on equivalence-checking of reversible circuits which appear in modules of quantum algorithms and require physical synthesis optimizations~\cite{FowlerEtAl2004} that must be verified.
However, other important modules in quantum algorithms,
such as the {\it Quantum Fourier Transform (QFT)},
are properly-quantum, and conventional circuits,
such as {\it modular exponentiation},
can be optimized for performance using properly-quantum gates.
Fortunately, simple cancellations in reversible miters can be used with properly-quantum circuits.
 Reduced properly-quantum miters can be verified using symbolic simulation with QuIDDPro~\cite{VMH03} or QMDD software~\cite{MT06}. Using reversible miters as pre-processors can dramatically decrease overall runtime.
 We empirically compare the following two methods.

\noindent{\bf With Local Simplification.}
Before invoking QuIDDPro, reduce the miter using local simplification.

\noindent{\bf  W/o Local Simplification.}
Apply QuIDDPro directly to the miter.

For properly-quantum circuit benchmarks, we used QFT and {\it modular
exponentiation} modules from circuits that implement Shor's factorization
algorithm on an LNN architecture~\cite{FowlerEtAl2004}.
For each benchmark circuit with $n$ inputs, we studied five cases (new gates were added in the middle).

\noindent{\bf Same.} Two identical copies of a benchmark circuit.

\noindent{\bf Different 1.}
A circuit and its copy with one gate added.

\noindent{\bf Different 2.}
A circuit and its copy with two gates added.

\noindent{\bf Different 3.}
A circuit and its copy
with one gate deleted.

\noindent{\bf Different 4.}
A circuit and its copy
with two gates deleted.

In Tables ~\ref{tab:exp5-2} to \ref{tab:exp5-3} we report
  runtimes for local simplification of reversible miters and
 subsequent QuIDDPro calls, subject to
 a $1000s$ time-out. In the
``Same'' case,
simplification alone proved equivalence.
However, in the ``Diff.~2'' case, many gates
remained after simplification and QuIDD runtimes
were substantial. In all cases, local simplification
improved overall runtimes.

\begin{table}[tp]
\begin{center}
\tcaption{Verification of QFT circuits without local simplification.}\label{tab:exp5-2}
\begin{tabular}{|c|c|c|c|c|c|c|c|c|c|c|} \hline
$n$ & \multicolumn{2}{|c|}{Same}   & \multicolumn{2}{|c|}{Diff. 1}   & \multicolumn{2}{|c|}{Diff. 2}   & \multicolumn{2}{|c|}{Diff. 3}   & \multicolumn{2}{|c|}{Diff. 4} \\ \cline{2-11}
& \multicolumn{2}{|c|}{simp. + QuIDD} & \multicolumn{2}{|c|}{simp. +  QuIDD} & \multicolumn{2}{|c|}{simp. + QuIDD} & \multicolumn{2}{|c|}{simp. + QuIDD} & \multicolumn{2}{|c|}{simp. + QuIDD} \\ \hline
 4 & \B - & \B  0.15
 & \B - & \B  0.15
 & - &   0.16
 & - &   0.14
 & - &   0.14
 \\  \hline
 8   & \B-  & \B  1.75
 & \B - & \B  1.80
 & - &   1.97
 & - &   1.74
 & - &   1.83
 \\  \hline
 16   & \B -  & \B   $>$ 1,000
 & \B - & \B   $>$ 1,000
 & - &    $>$ 1,000
 & - &    $>$ 1,000
 & - &    $>$ 1,000
 \\  \hline
 32   & \B - & \B   $>$ 1,000
 & \B - & \B   $>$ 1,000
 & - &    $>$ 1,000
 & - &    $>$ 1,000
 & - &    $>$ 1,000
 \\  \hline
 64   & \B - & \B   $>$ 1,000
 & \B - & \B   $>$ 1,000
 & - &    $>$ 1,000
 & - &    $>$ 1,000
 & - &    $>$ 1,000
 \\  \hline
\end{tabular}
\end{center}
\end{table}

\begin{table}[tp]
\begin{center}
\tcaption{Verification of QFT circuits with local simplification.}\label{tab:exp5-1}
\begin{tabular}{|c|c|c|c|c|c|c|c|c|c|c|} \hline
$n$ & \multicolumn{2}{|c|}{Same}   & \multicolumn{2}{|c|}{Diff. 1}   & \multicolumn{2}{|c|}{Diff. 2}   & \multicolumn{2}{|c|}{Diff. 3}   & \multicolumn{2}{|c|}{Diff. 4} \\ \cline{2-11}
& \multicolumn{2}{|c|}{simp. + QuIDD} & \multicolumn{2}{|c|}{simp. +  QuIDD} & \multicolumn{2}{|c|}{simp. + QuIDD} & \multicolumn{2}{|c|}{simp. + QuIDD} & \multicolumn{2}{|c|}{simp. + QuIDD} \\ \hline
 4   & \B 0        & \B  -  & \B  0 & \B   0.03     &   0 &    0.05  &   0 &   0.04  &  0 &   0.05  \\  \hline
 8   & \B  0       & \B -   & \B  0.01 & \B   0.03  &   0 &    0.17  &   0 &    0.04  &   0 &    0.26  \\  \hline
 16   & \B  0.05   & \B -  & \B  0.07 & \B   0.05   &   0.08 &    0.26  &   0.06 &    0.04  &   0.07 &    0.05  \\  \hline
 32   & \B  0.73   & \B -   & \B  1.11 & \B   0.04  &   1.13 &    9.17  &   0.99 &    0.04 &  1.22 &    0.08  \\  \hline
 64   & \B  17.29  & \B -   & \B  24.32 & \B   0.05 &   25.48 &    0.52 &   24.33 &   0.06 &   30.35 &    0.12 \\  \hline
 128   & \B  354.52 & \B -   & \B  366.2 & \B  0.04 &   497.21 &   $>$ 1,000  &   522.57 &    0.04 &   580.11 &    0.39 \\  \hline
\end{tabular}
\end{center}
\end{table}

\begin{table}[tp]
\begin{center}
\tcaption{Verification of modular multiplication w/o local simplification.}\label{tab:exp5-4}
\begin{tabular}{|c|c|c|c|c|c|c|c|c|c|c|} \hline
$n$  & \multicolumn{2}{|c|}{Same}   & \multicolumn{2}{|c|}{Diff. 1}   & \multicolumn{2}{|c|}{Diff. 2}   & \multicolumn{2}{|c|}{Diff. 3}   & \multicolumn{2}{|c|}{Diff. 4} \\ \cline{2-11}
& \multicolumn{2}{|c|}{simp. + QuIDD} & \multicolumn{2}{|c|}{simp. +  QuIDD} & \multicolumn{2}{|c|}{simp. + QuIDD} & \multicolumn{2}{|c|}{simp. + QuIDD} & \multicolumn{2}{|c|}{simp. + QuIDD} \\ \hline
 4 & \B - &  \B $>$ 1,000
 & \B - &  \B  $>$ 1,000
 & - &    $>$ 1,000
 & - &    $>$ 1,000
 & - &    $>$ 1,000
 \\  \hline
 8 & \B - &  \B  $>$ 1,000
 & \B - &  \B  $>$ 1,000
 & - &    $>$ 1,000
 & - &    $>$ 1,000
 & - &    $>$ 1,000
 \\  \hline
\end{tabular}
\end{center}
\end{table}

\begin{table}[tbp]
\begin{center}
\tcaption{Verification of modular multiplication with local simplification.}\label{tab:exp5-3}
\begin{tabular}{|c|c|c|c|c|c|c|c|c|c|c|} \hline
$n$  & \multicolumn{2}{|c|}{Same}   & \multicolumn{2}{|c|}{Diff. 1}   & \multicolumn{2}{|c|}{Diff. 2}   & \multicolumn{2}{|c|}{Diff. 3}   & \multicolumn{2}{|c|}{Diff. 4} \\ \cline{2-11}
& \multicolumn{2}{|c|}{simp. + QuIDD} & \multicolumn{2}{|c|}{simp. +  QuIDD} & \multicolumn{2}{|c|}{simp. + QuIDD} & \multicolumn{2}{|c|}{simp. + QuIDD} & \multicolumn{2}{|c|}{simp. + QuIDD} \\ \hline
 4 & \B 0.58 & \B -   & \B  0.98 & \B  0.04
 &  1.07 &   0.85
 &  0.98 &   0.05
 &  1.02 &   0.39
 \\  \hline
 8 & \B 2.13 & \B -   & \B 3.72 & \B  0.04
 &  3.69 &   0.37
 &  3.73 &   0.04
 &  3.45 &   1.19
 \\  \hline
 16 & \B 6.03 & \B -   & \B 10.11 & \B  0.05
 &  11.29 &    $>$ 1,000
 &  11.16 &   0.05
 &  11.26 &   5.73
 \\  \hline
 32 & \B 16.33 & \B -   & \B 27.65 & \B  0.04
 &  27.49 &   3.68
 &  27.21 &   0.05
 &  27.83 &   0.04
 \\  \hline
 64 & \B 36.28 & \B -   & \B 58.32 & \B  0.02
 &  59.27 &   0.56
 &  60.91 &   0.05
 &  60.13 &   1.33
 \\  \hline
 128 & \B 74.77 & \B -   & \B 119.71 & \B  0.04
 &  120.98 &   1.88
 &  120.83 &   0.05
 &  121.59 &   52.55
 \\  \hline
\end{tabular}
\end{center}
\end{table}

For a more convincing example, we check equivalence between
an LNN and non-LNN implementation (without {\it measurement gates})
of Shor's algorithm for factoring the number 15.
These equivalent properly-quantum circuits include 2,732 gates for the non-LNN version and 5,120 gates for the LNN version.
Their structure is very different. For equivalence-checking, we used QuIDDPro with and without local simplification, and
these runs completed in 59.07s and 64095.22s, resp.  The results
confirm
the effectiveness of local simplifications with
 reversible properly-quantum miters.

\subsubsection{Proposed Method: Boosting Verification by Using SAT-based Combinational Tools}

Local simplification may leave many gates around,
after which QuIDDPro tends to consume significant time and memory. However, if very few properly-quantum gates
remain, a more lightweight verification procedure may be used. Generic symbolic simulators, such as QuIDDPro, do not scale (empirically) as well as leading-edge SAT-based combinational equivalence-checking used in the Electronics industry to verify modern digital circuits (Sec.~\ref{sec:ConvVerify}).
Hence we leverage SAT-based tools to boost equivalence-checking of quantum circuits.

\vspace*{2mm}
\noindent
{\underline{\bf \sc For two circuits $C_1$ and $C_2$, we do the following}}.


\noindent{\bf Step 1.} Construct the miter circuit $C = C_1 \cdot C_2^{-1}$.

\noindent{\bf Step 2.} Perform simplification of the miter circuit.

\noindent{\bf  Step 3.} If properly-quantum gates remain, go to Step~4, else invoke state-of-the-art SAT-based combinational equivalence-checking (the ``cec'' command of ABC
  system~\cite{ABC}) to tell if the miter circuit is equivalent to Identity.

\noindent{\bf Step 4.} Find the longest sequence of conventional logic gates (NOT, CNOT, Toffoli) in the miter circuit. Label this sequence $C_a$. Let the simplified miter
circuit be  $Q_a \cdot C_a \cdot Q_b$.

\noindent{\bf Step 5.} Transform $Q_a \cdot C_a \cdot Q_b$ to
  $C_a \cdot Q_b \cdot Q_a$. Note that $Q_a \cdot C_a \cdot Q_b = I$ (Identity) iff $C_a \cdot Q_b \cdot Q_a
= I$ as shown in Sec.~\ref{sec:ReversibleMiterSimplification}.
  Move conventional gates in $Q_b \cdot Q_a$ to the front
  of the miter as much as possible, creating a transformed miter $C_a' \cdot Q_b'$, where  $C_a'$ and $Q_b'$ are a reversible circuit and a properly-quantum circuit, respectively.

\noindent{\bf Step 6.} Check the functionality of $Q_b'$ by lightweight iterated simulation. If it is not properly quantum, conclude that the miter circuit is not Identity. Else, go to Step 7.

\noindent{\bf Step 7.} Exploit the functionality of $Q_b'$, and let $C_b$ be
  a conventional circuit which corresponds to the exploited logic
  functionality. Then, check whether $C_a' \cdot C_b$ is Identity or
  not.


Suppose we have few properly-quantum gates as shown in the left-hand side of
Fig.~\ref{fig:extension} where $C_1$ is relatively large.
Then after Step~5, we can get the right-hand side circuit from the left-hand side circuit in Fig.~\ref{fig:extension}. Our miter becomes $C_1 \cdot Q_2$ where $C_1$ is reversible but $Q_2$ is properly-quantum. This avoids a heavy-duty generic quantum simulator for $C_1$.

A key observation is that the functionality of $Q_b'$ (at Step~6) should be
classical (inverse of $C_a'$) if
the entire miter is Identity. Thus, if $Q_b'$ is properly-quantum, the miter circuit is not Identity.
When $Q_b'$ has few gates, this can be checked efficiently by a quantum generic simulator. By Step~7, properly-quantum gates are reduced, and we can use state-of-the-art SAT-based combinational equivalence-checking.
By avoiding heavy-duty generic quantum simulation, our adaptive method can achieve significant speed-ups when $C_a'$ is large.

To validate our method, we studied circuits implementing
one iteration of Grover's quantum algorithm for
search~\cite{Grover1997}
 as shown in  Fig.~\ref{fig:GS}.
A particular step of the algorithm, called {\em the oracle},
is implemented with a reversible circuit module $C_f$ based
on a user-defined Boolean function $f$ (search predicate).
To make verification more challenging, we configured
a search predicate that contains a multiplier circuit.
We then created an equivalent variant of $C_f$ by
applying a global, rather than local, circuit transform.
Namely, we applied a certain wire permutation on inputs
of $C_f$ and its inverse on outputs of $C_0$. This permutation
was implemented by applying SWAP gates to (all) pairs
of adjacent wires and then breaking down each SWAP gate
into three CNOT gates, as described in Section \ref{sec:QC}.
In our case study, the proposed procedure goes as follows.

\noindent{\bf Step 1.} Construct the miter circuit $C = C_1 \cdot C_2^{-1}  =
C_f^{1} \cdot W^{1} \cdot C_0^{1} \cdot W^{1} \cdot
(W^{2})^{-1} \cdot (C_0^{2})^{-1} \cdot (W^{2})^{-1} \cdot
(C_f^{2})^{-1}$.

\noindent{\bf Step 2.} Simplify the miter circuit. Because of the inserted SWAP gates (if we use only naive cancellation rules), we cannot cancel the two pairs of $C_f^{1}$ and $(C_f^{2})^{-1}$, or $C_0^{1}$ and $(C_0^{2})^{-1}$.
But we can remove the sequence $W^{1} \cdot (W^{2})^{-1}$,
reducing the miter to
$C_f^{1} \cdot W^{1} \cdot C_0^{1} \cdot (C_0^{2})^{-1} \cdot
(W^{2})^{-1} \cdot (C_f^{2})^{-1}$.

\noindent{\bf Step 3.} Since properly-quantum gates remain, go to Step~4.

\noindent{\bf Steps 4 and 5.}
Move $(C_f^{2})^{-1}$ to the input side of the circuit
to maximize the conventional logic part in the prefix.
The miter becomes
$C_a' \cdot Q_b'$
where  $C_a' = (C_f^{2})^{-1} \cdot C_f^{1}$ and
$Q_b' =  W^{1} \cdot C_0^{1} \cdot (C_0^{2})^{-1} \cdot (W^{2})^{-1}$.

\noindent{\bf Steps 6. and 7.}
Using techniques described earlier, combine
 a quantum generic simulator (QuIDDPro~\cite{VMH03,VMH07})
and state-of-the-art SAT-based combinational equivalence-checking (the ``cec'' command of ABC system~\cite{ABC}).

The above technique is compared to constructing a miter circuit and applying the symbolic simulator QuIDDPro \cite{VMH03,VMH07} to the miter. QuIDDPro alone does not finish in ten hours,
but our technique completes in under seven seconds.

\section{Conclusion and Future Work}

We have studied several techniques for
equivalence-checking of reversible circuits, including the new
concept of reversible miters. In particular, we have observed that
state-of-the-art SAT-based combinational equivalence-checking (cec)
can be adapted to this context and outperforms generic quantum techniques.
Basic BDD-based techniques usually outperform SAT-based techniques,
but not cec. As is the case with ATPG, reversibility can significantly
simplify equivalence-checking, while these simplifications are
compatible with other techniques and amplify them.
We then proposed an adaptive method to verify quantum circuits more
efficiently than the existing quantum circuit verification tools by combining them
with the state-of-the-art SAT-based combinational
equivalence-checking tool for the conventional circuits.
Experiments suggest that reversible miters are
useful for the verification of reversible circuits as well as
properly-quantum circuits.




\end{document}